\documentclass[twocolumn,showpacs,pra,aps]{revtex4}
\usepackage{amssymb}
\usepackage{amsmath}
\usepackage{amsfonts}
\usepackage{graphics}
\usepackage{epic}
\usepackage{eepic}
\usepackage{color}
\usepackage{epsfig}
\usepackage{bbm}

\begin{document}

\def\ra{\rangle}
\def\la{\langle}
\def\bege{\begin{equation}}
\def\ende{\end{equation}}
\def\begarr{\begin{eqnarray}}
\def\endarr{\end{eqnarray}}
\def\ha{{\hat a}}
\def\hb{{\hat b}}
\def\hu{{\hat u}}
\def\hv{{\hat v}}
\def\hc{{\hat c}}
\def\hd{{\hat d}}
\def\no{\noindent}\def\non{\nonumber}
\def\hi{\hangindent=45pt}
\def\v{\vskip 12pt}

\newcommand{\bra}[1]{\left\langle #1 \right\vert}
\newcommand{\ket}[1]{\left\vert #1 \right\rangle}

\title{Towards photostatistics from photon-number discriminating detectors
}

\author{Hwang Lee$^1$}
\author{Ulvi H.\ Yurtsever$^1$}
\author{Pieter Kok$^{1,2}$} 
\author{George M.\ Hockney$^1$}
\author{Christoph Adami$^1$}
\author{Samuel L.\ Braunstein$^3$}
\author{Jonathan P.\ Dowling$^1$} 

\affiliation{
$^1$Quantum Computing Technologies Group, 
Section 367, Jet Propulsion Laboratory, \\
California Institute of Technology, MS 126-347,
 4800 Oak Grove Drive, CA~91109-8099 \\
$^2$Hewlett Packard Laboratories, Bristol  BS34 8QZ, United Kingdom \\ 
$^3$Computer Science, University of York, 
York YO10 5DD, United Kingdom
}

\date{\today}


\begin{abstract}
We study the properties of a photodetector that
has a number-resolving capability. In the absence of dark counts, 
due to its finite quantum efficiency, photodetection with such a detector
can only eliminate the possibility that the incident field
corresponds to a number of photons less than the detected photon number. 
We show that such a {\em non-photon} number-discriminating 
detector, however, provides a useful tool in the reconstruction of the
photon number distribution of the incident field even in the presence of dark counts.
\end{abstract}

\maketitle

\section{The Problem}

With the recent advent of linear optical quantum computation (LOQC), interest in
photon number-resolving detectors has been growing. The ability to 
discriminate the number of incoming photons plays an essential role in the
realization of nonlinear quantum gates 
in LOQC \cite{knill01,pittman02,franson02,ralph03}
as well as in quantum state  preparation \cite{lee02,kok02,fiurasek02,gerry02,pryde03}.
Such detectors are used to post-select particular quantum states of a superposition
and consequently produce the desired nonlinear interactions of LOQC.
In addition, the most probing attacks an eavesdropper can
launch against a typical quantum cryptography system exploit
photon-number resolving capability \cite{gilbert00,gisin02}.
Recent efforts in the development of such photon number-resolving
detectors include the visible light photon counter \cite{kim99,takeuchi99},
fiber-loop  detectors \cite{banaszek03,haderka03,rehacek03,achilles03,fitch03},
and superconducting transition edge sensors~\cite{miller03}.
Standard photodetectors can measure only the presence or absence of 
light (single-photon sensitivity), 
and generally do not have the capability of discriminating the number of incoming photons 
(single-photon number resolution). 
There have been suggestions of accomplishing
single-photon resolution using many single-photon--sensitive detectors arranged in a detector array or detector cascade \cite{song90,paul96,kok01,kok03}.
For example, the VLPC (visible light photon counter) is based on a confined avalanche breakdown in a small portion of the total detection area and, hence,
can be modeled as a detector cascade \cite{bartlett02}. 
Then again, fiber-loop detectors may be regarded as a detector cascade in the time domain.

Let us suppose a number-resolving detector
detects two photons in a given time interval.
If the quantum efficiency of the detector is one,
we can be certain that two photons came from the incident
light during that time interval. If the quantum efficiency is, say, 0.2,
what can we say about the incident light? 
In the absence of dark counts, we can safely say only that 
there were {\em at least} two photons in the incident beam. 
In this case, photodetection rather has a 
{\em non-photon} number discriminating feature,
since the {\em conditional} probability of {\em not} having zero,
or one photon in the incident pulse is zero. Can we say more than that?
For example, what is the probability that the incident pulse 
actually corresponds to two photons, or three? 
In this paper we attempt to answer these questions and
discuss the effect of quantum efficiency on the photon-number resolving capability.

The semiclassical treatment of photon counting statistics  was first derived by Mandel \cite{mandel58}. He showed that  the photon counting distribution is Poissonian
if the intensity of the incident light beam is constant in time.
The full quantum mechanical description has been
formulated by Kelley and Kleiner~\cite{kelley64}, 
and Glauber \cite{glauber65}. Here, we start from a formula based on the 
photon-number state representation, which was first
derived by Scully and Lamb \cite{scully69}.
For a given quantum efficiency of the photodetector, 
the photon counting distribution can be written as
\begarr
P(k) = \sum_{n=k}^\infty {n \choose k} \eta^k (1-\eta)^{n-k} S(n) ,
\label{prob}
\endarr

\no
where $P(k)$ is the probability of detecting $k$ photons,
$\eta$ is the quantum efficiency, 
and $S(n)$ is the probability that the source (incident light)
corresponded to $n$ photons.
It is assumed that the quantum efficiency $\eta$ 
does not depend on the intensity of light. 
This distribution then leads to a natural definition of  $P(k|n)$
as the conditional probability of detecting $k$ photons given that
$n$ photons being in the source as
\begarr
P(k|n) = {n \choose k} \eta^k (1-\eta)^{n-k}.
\label{condi-1}
\endarr

Contrariwise, suppose we detect $k$ photons.
If the detector has a quantum efficiency much less than unity, 
very little can be said about the actual photon number
associated with the source. 
We may, however, deduce that the source actually corresponded to
$n$ photons with a certain probability. 
This probability should be high if the quantum efficiency is high. 
In the following, we will use Bayes's theorem in order to extract information about the source from the conditional photon-counting statistics. 
Armed with  the conditional probability $P(k|n)$, we may write

\begarr
Q(n|k)
=
{ P(k|n) S(n) \over P(k) } 
=
{ P(k|n) S(n) \over \sum_i P(k|i) S(i) } 
.
\label{condi-2}
\endarr

Now we will identify the conditional probability, 
$Q(n|k)$ with the probability that an {\em ideal detector} 
would have detected $n$ photons,
given that $k$ photons are detected by the imperfect detector. 
It is given  as 
the product of  the probability of the source corresponding to $n$ photons 
{\em before the measurement} and
the probability of $k$-photon detection, given that $n$ photons are in the
source, normalized to the probability of having 
$k$ photons detected. 

It is evident that generally we cannot find $Q(n|k)$
if we do not know $S(n)$--the distribution of the source.
The exceptions are: (i) If $n < k$, then $P(k|n) =0$.
Therefore, $Q(n|k) =0$ for $n < k$ no matter what $S(n)$ is.
(ii) If the quantum efficiency is unity, that is $\eta=1$,
then $P(k|n) = 0$ when $n \neq k$, and
$P(k|n=k)=1$. Since $P(k) = S(n=k)$, it leads to
$Q(n=k|k)=1$ and $Q(n\neq k|k)=0$.
Hence, without knowing $S(n)$, we only have information about $Q(n|k)$ 
when $n < k$ or $\eta =1$.
Such a detector, then, readily {\em excludes} certain numbers of photons
in the source rather than determining them. 
Without a prior knowledge of the source, if $\eta \neq 1$, 
we call this a {\em non-photon number-discriminating detector}.

\begin{figure}[b]
\centerline{\psfig{figure=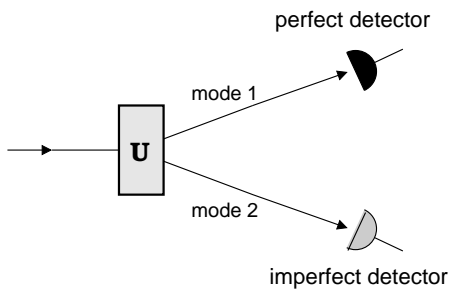,width=6cm}}
\bigskip
\caption{\label{heralded} Schematic of the heralded-photon setup.  
   }
\end{figure}

Obviously, the difficulty in talking about
this conditional probability
lies in the fact that the photodetection is destructive.
We will elaborate this subtlety
using a heralded-photon setup
depicted in Fig.~\ref{heralded}.
The unitary operator ${\hat U}_H$
transforms the input state in the following way:
\begarr
{\hat U}_H \sum_{n=0}^\infty \alpha_n |n\ra
= \sum_{n=0}^\infty \alpha_n |n\ra_1 |n\ra_2 .
\endarr

\no
Suppose now the photon number in the mode 1
is measured by the perfect detector and 
the photon number in the mode 2
is measured by the imperfect detector.
It is now simple to understand the conditional
probability of finding $n$ photons in the mode 1, given that
$k$ photons are detected in the mode 2.
In this way, we can have an operational definition
of the conditional probability $Q(n|k)$. It is easy to show that
the conditional probability of finding $n$ photons in mode 1
given that $k$ photons are detected in mode 2 is given by precisely
the same expression for the quantity $Q(n,k)$ as Eq.\,(3).
[Note that
the transformation given by ${\hat U}_H$ is far from quantum cloning,
since the result of $U$ is far from being
equal to the cloning transformation
$(\sum_{n=0}^\infty \alpha_n |n\ra_1) \otimes
(\sum_{n=0}^\infty \alpha_n |n\ra_2)$.

\section{POVM Description}

Suppose an initial quantum state $|\psi_{in}\ra$ of incident light
is represented in the photon number states, as
\begarr
|\psi_{in}\ra = \sum_{n=0}^\infty \alpha_n |n\ra .
\endarr
\no
The probability of finding $n$ photons 
is then given by $S(n)= |\la n |\psi \ra|^2=|\alpha_n|^2$.
We may describe the detection process of the perfect detector 
by entangling the source with the detector ancilla $|0\ra$
\begarr
|\psi_{f}\ra = {\hat U}| \psi_{in} \ra |0\ra = \sum_n \alpha_n |n\ra_Q |a_n\ra_M\;,
\endarr
where
\begarr
{\hat U} = \sum_j |j\ra \la j| \otimes \frac1{\sqrt{j!}}(\ha^\dagger)^j 
\label{vNperfect}
\endarr
is the unitary operator implementing the von Neumann measurement, 
and we identified the quantum ($Q$) and measurement device $(M)$ Hilbert spaces 
with appropriate subscripts. 

The probability to detect $k$ photons, $P(k)$, can now be calculated as
\begarr
P(k) = {\rm Tr}_Q (|a_k\ra_M\la a_k| |\psi_{f}\ra \la \psi_{fi}|) = |\alpha_k|^2\;,
\label{p-k-1}
\endarr 
as expected. 
Of course, the same result would have been obtained as  $P(k) ={\rm Tr}(M_k\rho_{in})$ using a Positive Operator Valued Operator Measure (POVM) 
\begarr
M_k = |k\ra \la k|\;.
\endarr
The present unitary treatment, however, is more transparent and 
allows a straightforward extension to imperfect detectors.

For an imperfect photodetector with a finite quantum efficiency,  
the POVM can be written as 
\begarr
M_k = \sum_{n=k}^\infty P(k|n) |a_n\ra \la a_n|, \label{povm}
\endarr

\no
with $P(k|n)$ given by Eq.~(\ref{condi-1}). 
We can verify this by calculating as before
\begarr
|\psi_{f}\ra = U| \psi_{in}\ra |0\ra_M |0\ra_E \;,
\endarr
except that the entanglement operator $U$ now acts on the joint Hilbert space of the quantum system, a measurement device, and an environment $|0\ra_E$, which is where our lost photons go. Thus, the unitary operator can be given by
\begarr
{\hat U} = \sum_j |j\ra \la j| \otimes \frac1{\sqrt{j!}}\left(\sqrt{\eta}\, \ha^\dagger \otimes \mathbbm{1} + e^{i \varphi} \sqrt{1-\eta} \, \mathbbm{1} \otimes \hb^\dagger \right)^j\;,
\non
\endarr
where
${\varphi}$ is an arbitrary phase and 
$b^\dagger$ is the creation operator for the relevant
mode of the environment.
Similar to Eq.~(\ref{p-k-1}), we now have 
\begarr
P(k)&  =& {\rm Tr}_{QE} \left(|a_k\ra_M \la a_k||\psi_{fi}\ra \la \psi_{fi}|\right) 
\nonumber \\
 & =&  \sum_{n=k}^\infty |\alpha_n|^2 {n \choose k} \eta^k (1-\eta)^{n-k} \;.
\endarr
supporting our Eq.(\ref{povm}) as the correct POVM. 
Note that any relative phase information between the photon number states
disappeared. 

The POVM effectively transfers the initial quantum state of light 
into a certain quantum state
(as a result of $k$-photon detection) by the transformation
\begarr
\rho_{f-k} \equiv M_k \rho_{in} \;,
\endarr
\no
where $\rho_{in} = |\psi_{in}\ra \la \psi_{in}|$.
The probability that $k$ photons are detected can then also be written as
\begarr
P(k) &=& {\rm Tr} \left(\rho_{f-k} \right) \non \\
&=& {\rm Tr} \left(M_k |\psi_{in}\ra \la \psi_{in}| \right) \non \\
&=& \sum_n P(k|n) S(n).
\endarr
 
\no
On the other hand, the joint probability that the source state is $|n\ra$
and $k$ photons are detected
can be expressed by the probability 
\begarr
P(n,k) &=& {\rm Tr}_E(\la n, a_k | \psi_{f}\ra \la \psi_{f} |n, a_k\ra) \nonumber \\
&=&\la n|\rho_{f-k}|n\ra
=P(k|n) S(n) \;,
\endarr
which defines the quantity
$Q(n|k)$ as
\begarr
Q(n|k) \equiv
{\la a_n |\rho_{f-k} |a_n \ra
\over
{\rm Tr}\left(\rho_{f-k} \right)
}\;.
\endarr

In particular, the quantity $Q(n=k|k)$ corresponds to
the {\em confidence} of the state preparation in
a situation that
we have more than two modes for the input
and the photon number in one of the modes is
measured \cite{kok01}.

\section{An Example}


Let us take a simple example of an input coherent state
in order to see how the value of $Q(n|k)$ behaves.
In doing this, we assume that the dark count rate is
small enough that we can ignore it, 
and the quantum efficiency is independent of
the number of incident photons.

A coherent state $|\alpha\ra$ is expressed in terms of number state 
as
$|\alpha\ra = e^{-|\alpha|^2/2} \sum_n (\alpha^n / \sqrt{n!}) |n\ra
$ \cite{scully97}. 
The probability of finding $n$ photons in $|\alpha\ra$
is then given by a Poisson distribution:
\begarr
S(n) = |\la n|\alpha\ra|^2
={e^{-{\bar n}}  {\bar n}^{n} \over n! } ,
\endarr

\no
where ${\bar n}= |\alpha|^2$.
Now the probability of detection of $k$ photons with
an imperfect detector is written as \cite{scully69}
\begarr
P(k) &=& \sum_{n=k}^{\infty} 
{n \choose k} \eta^k (1-\eta)^{n-k} 
{e^{-{\bar n}}  {\bar n}^{n} \over n! } 
\non \\
&=&
e^{-{\bar n}}  
\sum_{l=0}^{\infty} 
{{l+k} \choose k}  {{\bar n}^{l} {\bar n}^{k} \over (l+k)!} 
\eta^k (1-\eta)^{l} 
\non \\
&=&
e^{-{\bar n}} { {\bar n}^{k} \eta^k \over k!} 
\sum_{l=0}^{\infty} 
{{\bar n}^{l}  \over l!} 
(1-\eta)^{l} 
\non \\
&=&
e^{-{\bar n}} { ({\bar n} \eta)^{k} \over k!} 
e^{-{\bar n}(1-\eta)} 
=
{ ({\bar n} \eta)^{k} \over k!} 
e^{-{\bar n}\eta} .
\endarr

\no
That is, the photon counting
statistics are simply a Poisson distribution
where the average number of photons detected is
the average number of incident photons
multiplied by the quantum efficiency of the 
detector--given by ${\bar k}= {\bar n} \eta$ as expected.

\begin{figure}[b]
\centerline{\psfig{figure=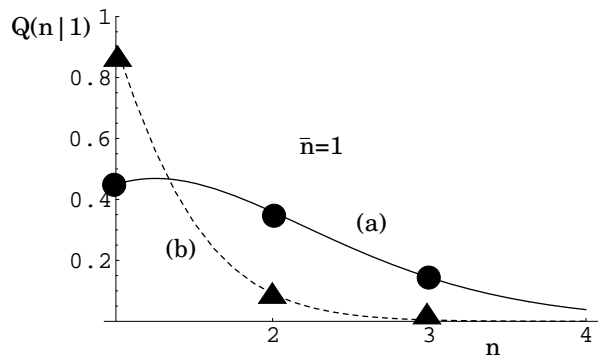,width=8cm}}
\bigskip
\caption{\label{fig:nbar=1} The conditional probability 
    $Q(n|k=1)$ for a coherent state with ${\bar n}=1$
    as a function of $n$. 
    (a) For detector efficiency $\eta=0.2$,
     $Q(1|1)=0.45$, $Q(2|1)=0.36$.
     (b) For detector efficiency $\eta=0.9$,
      $Q(1|1)=0.90$, $Q(2|1)=0.09$. 
   }
\label{n=1}
\end{figure}

\begin{figure}[t]
\centerline{\psfig{figure=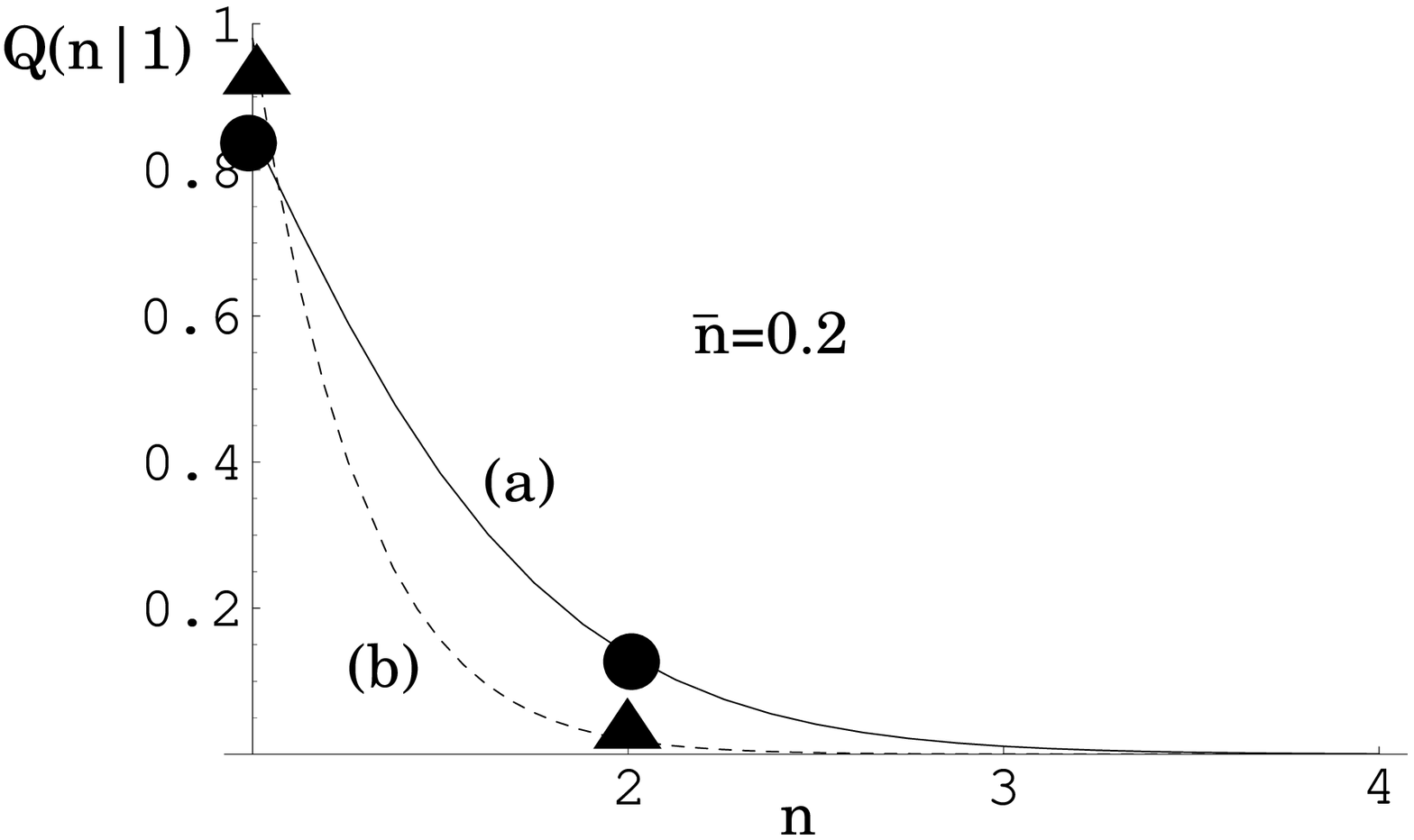,width=8cm}}
\bigskip
\caption{\label{fig:nbar=0.2} The conditional probability 
    $Q(n|k=1)$ for a coherent state with ${\bar n}=0.2$
    as a function of $n$. 
    (a) For detector efficiency $\eta=0.2$,
     $Q(1|1)=0.85$, $Q(2|1)=0.14$.
     (b) For detector efficiency $\eta=0.9$,
      $Q(1|1)=0.98$, $Q(2|1)=0.02$. 
   }
\label{n=0.2}
\end{figure}

Now let us consider the conditional probability $Q(n|k)$.
Using Eq.~(\ref{condi-2}), it is simply found as
\begarr
Q(n|k) &=&{
{n \choose k} \eta^k (1-\eta)^{n-k}
e^{-{\bar n}}  {\bar n}^n / n! \over 
e^{-{\bar n} \eta }  ({\bar n} \eta)^k / k! }
\non \\
&=&
{1 \over (n-k)!} (1-\eta)^{n-k} 
e^{-{\bar n}(1-\eta)} 
{\bar n}^{n-k}
\non \\
&=&
{e^{-{\bar l}} \over l! }
{\bar l}^{~l} ,
\endarr

\no
where $l=n-k$ and ${\bar l} ={\bar n} (1-\eta)$.
Therefore, the form of $Q(n|k)$ is again a Poisson distribution
as a function of $n-k$ with the average number of
photons that {\em fail to register}
given as ${\bar n} (1-\eta)$.

Figure \ref{n=1} shows the behavior of $Q(n|k)$ 
as a function of $n$, for the case of coherent state 
input with ${\bar n}=1$, when one photon is detected.
We can see that
if the quantum efficiency of the detector is
0.2, with the detection of one photon, then the
chances are more than 50\% that 
the input contains more than two photons ($Q(1|1)=0.45$).
Obviously, $Q(1|1)$ increases for higher values
of the quantum efficiency ($Q(1|1)=0.9$ for $\eta=0.9$).

On the other hand, for a given quantum efficiency of the 
detector, $Q(1|1)$ gets large as the average number of
photons in the input decreases.
For ${\bar n}=0.2$, $Q(1|1)=0.85$ even if the detector
efficiency is 0.2, as depicted in Fig.~\ref{n=0.2}.
In this case, with the detection of one photon 
the chances are 85\% that the detection result corresponds 
to the correct value.

\section{Reconstruction of Photon-Number Distribution}

So far we have seen that
we cannot say much about
$Q(n|k)$ if we do not know at least the form of $S(n)$.
In other words, using an imperfect detector,
even after a detection of $k$ photons, we do not gain
much information about the number of incident photons
without a prior knowledge of the incident field.
Hence, the problem comes back to
reconstruction of photon number distribution of
the incident field, $S(n)$, from 
the photon counting probability, $P(k)$ by
a sequence of identical inputs and measurements.
For this purpose, we may write Eq.~(\ref{condi-1})
in a matrix form as
\begarr
{\vec P} = {\bf P} {\vec S},
\endarr
\no
where
\begarr
{\vec P} &=&  \left( \begin{array}{c} 
P(0) \\ P(1) \\ P(2) \\ \cdot  \\ \cdot
\end{array}
\right), 
\quad
{\vec S}=  \left( \begin{array}{c} 
S(0) \\ S(1) \\ S(2) \\ \cdot  \\ \cdot
\end{array}
\right) 
\endarr
\no
and
each component of the matrix ${\bf P}$
represents the conditional probability $P(k|n)$
as 
\begarr
P_{kn} \equiv P(k|n)
= { n \choose k} \eta^k (1-\eta)^{n-k}.
\label{p-kn}
\endarr
\no
Then the inverse of ${\bf P}$ can be found
explicitly; it is given by
\begarr
(P^{-1})_{nm} =
{ m \choose n} \eta^{-m} (\eta -1)^{m-n} \; ,
\label{inv-1}
\endarr
\no
which, in the absence of dark counts,
determines the photon-number distribution of the source
from the photon-counting distribution.
Note, however, that for a fixed $n$
there are alternating signs in
$(P^{-1})_{mn}$ as the index $m$ changes,
so that this inversion formula can yield
unphysical results depending on the
probability distribution $P(k)$. 
Put another way,
for small efficiencies the matrix $\bf P$ is ill-conditioned
and ${\bf P}^{-1}$ will contain matrix elements that can get
very large in absolute value and can have either sign.
This means that
even if the calculated $S(n)$'s are 
all non-negative, a good inversion is
still not guaranteed, since
the inversion formula (\ref{inv-1}) is 
highly sensitive to small changes 
of $P(k)$, amplifying the statistical and other noise which is
inevitably present in any actual measurement.
The fact that the inverse matrix $P^{-1}$ is also
a upper triangular matrix implies that
$S(n)$ is determined by
$P(k)$ with $k\geq n$.
In other words, the tail of the measurement result, $P(k)$,
where the statistical noise can be expected to be greatest,
affects the inversion significantly. 
The validity conditions for the inversion formula 
shall be discussed elsewhere.

\begin{figure}[t]
\centerline{\psfig{figure=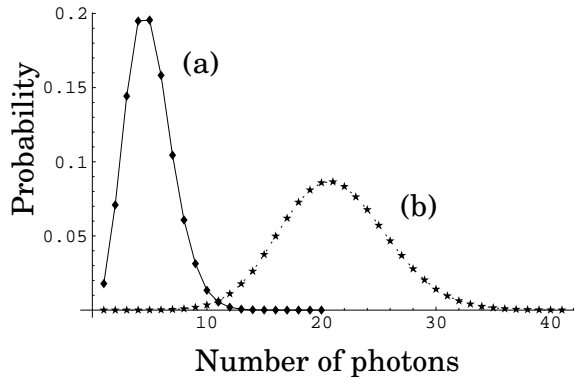,width=7.5cm}}
\bigskip
\caption{Reconstruction of the input-state
photon-number distribution:
(a) Photon counting distribution from the NIST experiment \cite{miller03};
(b) Reconstruction of the input state photon distribution
 (${\bar n} \approx 20.15$), assuming a quantum 
efficiency of 0.2 and no dark counts.
}
\label{fit}
\end{figure}

Instead, we can carry out the following recipe:
1) Seek an underlying distribution ${\vec S}$ for which 
${\bf P} {\vec S}$ matches
the observed frequencies according to a $\chi^2$ goodness-of-fit test.
2) Then, of all such distributions, seek the one with maximum entropy; 
{\em i.e.}, pick the most likely underlying distribution, 
which is statistically consistent with the data.
Such a strategy can be implemented as a genetic algorithm,
which we will discuss in a forthcoming paper \cite{yurtsever03}.
Using this method, the reconstruction of ${\vec S}$ with
a data set from an actual experiment is depicted in Fig.~\ref{fit}.
The plot in Fig.~\ref{fit}a shows the probability distribution of the
detected photons.
The raw count data was normalized by the total number of counts
(data provided courtesy of NIST \cite{miller03}.
The result of reconstruction 
is shown by the plot Fig.~\ref{fit}b,
and yield a ${\bar n}=20.15$ for the incident coherent state--a good
agreement with the experimental value of ${\bar n}_{\rm exp}=4.02$ \cite{miller03}.

\section{Dark Counts}

In addition to the finite quantum efficiency,
the dark counts further make the
results extracted from the experiment obscure.
For a constant dark-count rate, however, a generalization
of the inversion may be handled analytically.
Suppose in general that the probability of $d$ dark counts
is $D(d)$, a discrete probability distribution which
satisfies the normalization condition
\bege
\sum_{d=0}^{\infty} D(d) = 1 \; .
\ende

\no 
We assume that the distribution $D(d)$ represents probabilities
of dark counts in a fixed interval of time $\tau$ (duration of a
single photon-counting experiment), and that these probabilities
are independent of all other relevant observables such as incoming
photon number, number of detected photons, etc.
Therefore in the presence of dark counts, 
the conditional probability of detecting $k$ photons, 
given $n$ photons in the incident field Eq.~(\ref{condi-1}), is
modified accordingly:
\bege
P_D (k|n)
= \sum_{d=0}^k  D(d) 
{n \choose k-d} \eta^{k-d} (1-\eta)^{n-k+d} \;,
\label{pd-kn-1}
\ende

\no which expresses the probability $P_D (k|n)$ as a sum
(over all possible $d$) of the probability
of registering $k-d$ photons from the incident field
via true detection while
the remaining $d$ are registered via dark-count events.

As long as the physical time scale for
the emergence of a dark-count event is much
smaller than $\tau$, and the events arrive at a constant rate $\lambda$,
the probability distribution $D(d)$
(of the arrival of $d$ dark-count events in the fixed time
interval $\tau$) has a
Poissonian-distribution limit:
\bege
D(d) = e^{- \lambda \tau} \frac{(\lambda \tau )^d}{d!} \; ,
\label{d-d}
\ende

\no 
where the average number of dark count is
${\bar d} = \lambda \tau$.
Since after interchanging the summation index $d$
with $k-d$ we can rewrite Eq.~(\ref{pd-kn-1}) as
\bege
P_D (k|n)
= \sum_{d=0}^k  D(k-d) 
{n \choose d} \eta^{d} (1-\eta)^{n-d} \; ,
\label{condi-3}
\ende

\no 
where the summation over $d$
is in the form of a matrix multiplication
[Eq.\,(\ref{p-kn})],
\bege
( P_D )_{kn} = \sum_d D_{kd} \; P_{dn} ,\;
\ende

\no 
or in matrix notation,
\bege
{\bf P_D} = {\bf D} \, {\bf P} \; ,
\label{mat-pd}
\ende

\no 
where $(P_D )_{kn} \equiv P_D(k|n)$, and
\begarr
D_{kd} & \equiv & D(k-d)
\non \\
&=& e^{- \lambda \tau} \frac{(\lambda \tau )^{k-d}}{(k-d)!} \; .
\label{D-kd}
\endarr

\no 
Throughout the paper, we adopt the convention
that $k! = \Gamma(k+1) \equiv \infty $ for negative integer $k$, 
and correspondingly ${a \choose b} \equiv 0 $ whenever $b > a$.
In Eq.~(\ref{D-kd}), this convention implies that $D_{kd} = 0 $ for $d > k$.

Now by inverting Eq.~(\ref{condi-3})
in a matrix form, from the fundamental equation,
\bege
P(k) = \sum_n P_D(k|n) \, S(n) \;,
\label{p-k-2}
\ende

\no 
we can reconstruct the photon number
distribution $S(n)$ of the incident field given
the vector of observed photon-count
probabilities $P(k)$, whose relationship
to $S(n)$ [given by Eq.~(\ref{p-k-2})], can again be written in
matrix form as $\vec{P} = {\bf P_D } \vec{S}$,
just as in Eqs.\,(20)--(21).
The inverse of the coefficient matrix $P_D(k|n) = (P_D)_{kn}$ is found
from Eq.\,(\ref{mat-pd}),
\bege
{\bf P_D}^{-1} = {\bf P}^{-1} \, {\bf D}^{-1} \; .
\ende

\no Let us first consider the
inverse of the matrix ${\bf D}$ defined as in Eq.\,(30).
It is not difficult to prove
that the inverse of ${\bf D}$ is given by
\begarr
(D^{-1})_{ij} 
&=&
e^{\lambda \tau} \frac{(-\lambda \tau)^{i-j}}{(i-j)!} \non \\
& = & e^{\lambda \tau} \,
(-1)^{i-j} \, \frac{(\lambda \tau)^{i-j}}{(i-j)!}\; .
\label{inv-2}
\endarr

\no 
Combining Eq.\,(33) with Eq.\,(32) and Eq.\,(23),
we find
\begarr
({P_D}^{-1})_{kn}
& = & e^{\lambda \tau}
\sum_i {i \choose k} \eta^{-i} (\eta -1)^{i-k}
\frac{(-\lambda \tau)^{i-n}}{(i-n)!} \non \\
& = & \frac{(-1)^{k+n} e^{\lambda \tau}}
{(1- \eta )^k (\lambda \tau)^n }
\sum_{i=0}^{\infty} 
\frac{{i \choose k}}{(i-n)!}
\left[ \frac{(1-\eta) \lambda \tau}{\eta} \right]^i
\; . \non \\
\endarr

\no 
The infinite power series over $i$ in Eq.\,(34) is absolutely
convergent for all values of the argument
\bege
z \equiv \frac{(1-\eta) \lambda \tau}{\eta} \; ,
\ende
\no 
and can be expressed in terms of the generalized
hyper-geometric function $_1 \! F_1 (a,b,z)$ (which is an entire
function of the argument $z$ for all parameter values $a$, $b$),
giving a compact expression
for the inverse matrix elements in the form:
\begarr
({P_D}^{-1})_{kn}  = (-1)^{k+n}
\frac{e^{\lambda \tau}}{(1-\eta )^k (\lambda \tau )^{n}} \times
\non \\
\left\{ \begin{array}{ll}
\frac{n!}{k! (n-k)!} \; z^n \; \, {_1 \! F_1} (n+1, \, n-k+1, \,
z) & \; \; \textrm{if $k \leqslant n \; ,$}\\
\\
\frac{\, 1 \,}{\, \; (k-n)! \; \,} \; z^k\; \, {_1 \! F_1} (k+1, \, k-n+1, \,
z) & \; \; \textrm{if $k > n \; .$}
\end{array} \right. \non \\
\; 
\endarr
The inversion of Eq.\,(32) can now be effected by substituting
Eqs.\,(35)--(36) into the matrix formula
\bege
S(k) = \sum_n ({P_D}^{-1})_{kn} \, P(n) \; .
\ende

\no
This suggests that the reconstruction of the photon statistics
of the source may be possible with any given quantum efficiency
of the detector, even in presence of dark counts \cite{yurtsever03}.
However, as the efficiency falls and the dark-count rate grows,
more and more data points will be required for the inversion.


\section{Summary}

We have shown a way of quantifying the capability of
photon-number discriminating detectors by the conditional
probability, $Q(n|k)$, given in Eq.~(\ref{condi-2}).
This conditional probability, however,
can be obtained only when a priori knowledge of
the input is available.
Without a prior knowledge of the input state the results of
photodetection can only eliminate the possibility of having 
certain numbers of photons
in the source rather than determine them
in the absence of dark counts. 
Hence, the term {\em non-photon number-discriminating detector}
corresponds to the situation where the input
state is unknown.
On the other hand, the same photodetector
could be a {\em photon number-discriminating detector}
if the input state is known.
Of course, the number resolving capability, without a prior
knowledge of the input, can be
measured qualitatively using the conditional probability
$P(k|n)$ of Eq.~(\ref{condi-1}), 
which is given by the characteristics of the detector itself.
In particular, $C_k\equiv P(k|K)$
may serve the measure of that capability.
For the detector model considered in this paper, 
the number-resolving capability up to two photons
can be written as $C_{k=2}=\eta^2$, and for $C_{2}$ to be bigger than
0.5, $\eta$ should be bigger than $1/\sqrt[2]{2} \approx 0.71$.
Similarly, $P_D(k|k)$ of Eq.~(\ref{condi-3}) plays the same role in the presence
of dark counts.
For example, with a given dark-count rate $\lambda \tau =0.5$,
the detector efficiency $\eta$ should be more than 0.78
for $P_D(k|k)$ to be larger than one half.

We then introduced a simple matrix inversion formula 
in order to reconstruct
the input-state photon distribution from the measured
photon counting distribution.
Due to the form of the inverse matrix and its
high sensitivity to the fluctuation,
restrictions need to be imposed
on the use of this inversion formalism.
Alternatively, we can use a genetic algorithm that
finds the distribution with maximum
entropy after passing the $\chi^2$ goodness-of-fit test.
We have shown good results for the reconstruction
of the input distribution using the data from 
a recent experiment at NIST.
Finally, we have considered the reconstruction task 
in presence of dark counts and derived
a completely analytic formula for the inversion.
A detailed analysis of the 
conditions for the inversion formula to be
applicable will be
discussed in future work.

\section*{Acknowledgment}

This work was carried out at the Jet Propulsion Laboratory, 
California Institute of Technology, under a contract with 
the National Aeronautics and Space Administration. 
We wish to thank 
G.S.\ Agarwal, J.D.\ Franson, and T.B.\ Pittman 
for helpful discussions as well as
A.J.\ Miller for sharing data from the experiment
performed at NIST.
We would like to acknowledge support from 
Advanced Research and Development Activity,
the Army Research Office, 
Defense Advanced Research Projects Agency, 
National Reconnaissance Office, 
National Security Agency, 
the Office of Naval Research,
and NASA Code Y.

\end{document}